\documentclass[aps,prb,reprint,superscriptaddress]{revtex4-2}

\usepackage{graphicx}
\usepackage{dcolumn}
\usepackage{bm}
\usepackage{amsmath}
\usepackage{amssymb}
\usepackage{hyperref}
\usepackage{xcolor}
\usepackage{gensymb}

\usepackage{pdfpages}
\usepackage{pgffor}
\makeatletter
\AtBeginDocument{\let\LS@rot\@undefined}
\makeatother

\begin{document}
	
	
	\title{Counterintuitive gate dependence of weak antilocalization in bilayer graphene/WSe$_2$ heterostructures}
	\author{Julia Amann}
	\affiliation{Institut f\"ur Experimentelle und Angewandte Physik, Universit\"at Regensburg, 93040 Regensburg, Germany}
	\author{Tobias V\"olkl}
	\altaffiliation[Present address: ]{Department of Condensed Matter Physics, Weizmann Institute of Science, Rehovot, Israel}
	\affiliation{Institut f\"ur Experimentelle und Angewandte Physik, Universit\"at Regensburg, 93040 Regensburg, Germany}
	\author{Tobias Rockinger}
	\affiliation{Institut f\"ur Experimentelle und Angewandte Physik, Universit\"at Regensburg, 93040 Regensburg, Germany}
	\author{Denis Kochan}
	\affiliation{Institut f\"ur Theoretische Physik, Universit\"at Regensburg, 93040 Regensburg, Germany}
	\author{Kenji Watanabe}
	\affiliation{Research Center for Functional Materials, 
		National Institute for Materials Science, 1-1 Namiki, Tsukuba 305-0044, Japan}
	\author{Takashi Taniguchi}
	\affiliation{International Center for Materials Nanoarchitectonics, 
		National Institute for Materials Science,  1-1 Namiki, Tsukuba 305-0044, Japan}
	\author{Jaroslav Fabian}
	\affiliation{Institut f\"ur Theoretische Physik, Universit\"at Regensburg, 93040 Regensburg, Germany}
	\author{Dieter Weiss}
	\affiliation{Institut f\"ur Experimentelle und Angewandte Physik, Universit\"at Regensburg, 93040 Regensburg, Germany}
	\author{Jonathan Eroms}
	\email{jonathan.eroms@ur.de}
	\affiliation{Institut f\"ur Experimentelle und Angewandte Physik, Universit\"at Regensburg, 93040 Regensburg, Germany}
	
	


	
	\date{\today}

\begin{abstract}
Strong gate control of proximity-induced spin-orbit coupling was recently predicted in bilayer graphene/transition metal dichalcogenides (BLG/TMDC) heterostructures, as charge carriers can easily be shifted between the two graphene layers, and only one of them is in close contact to the TMDC. The presence of spin-orbit coupling can be probed by weak antilocalization (WAL) in low field magnetotransport measurements. When the spin-orbit splitting in such a heterostructure increases with the out of plane electric displacement field $\bar D$, one intuitively expects a concomitant increase of WAL visibility. Our experiments show that this is not the case. Instead, we observe a maximum of WAL visibility around $\bar D=0$. This counterintuitive behaviour originates in the intricate dependence of WAL in graphene on symmetric and antisymmetric spin lifetimes, caused by the valley-Zeeman and Rashba terms, respectively. Our observations are confirmed by calculating spin precession and spin lifetimes from an $8\times 8$ model Hamiltonian of BLG/TMDC.  
\end{abstract}

\maketitle

\section{Introduction}

\begin{figure}
	\includegraphics[width=8.6cm]{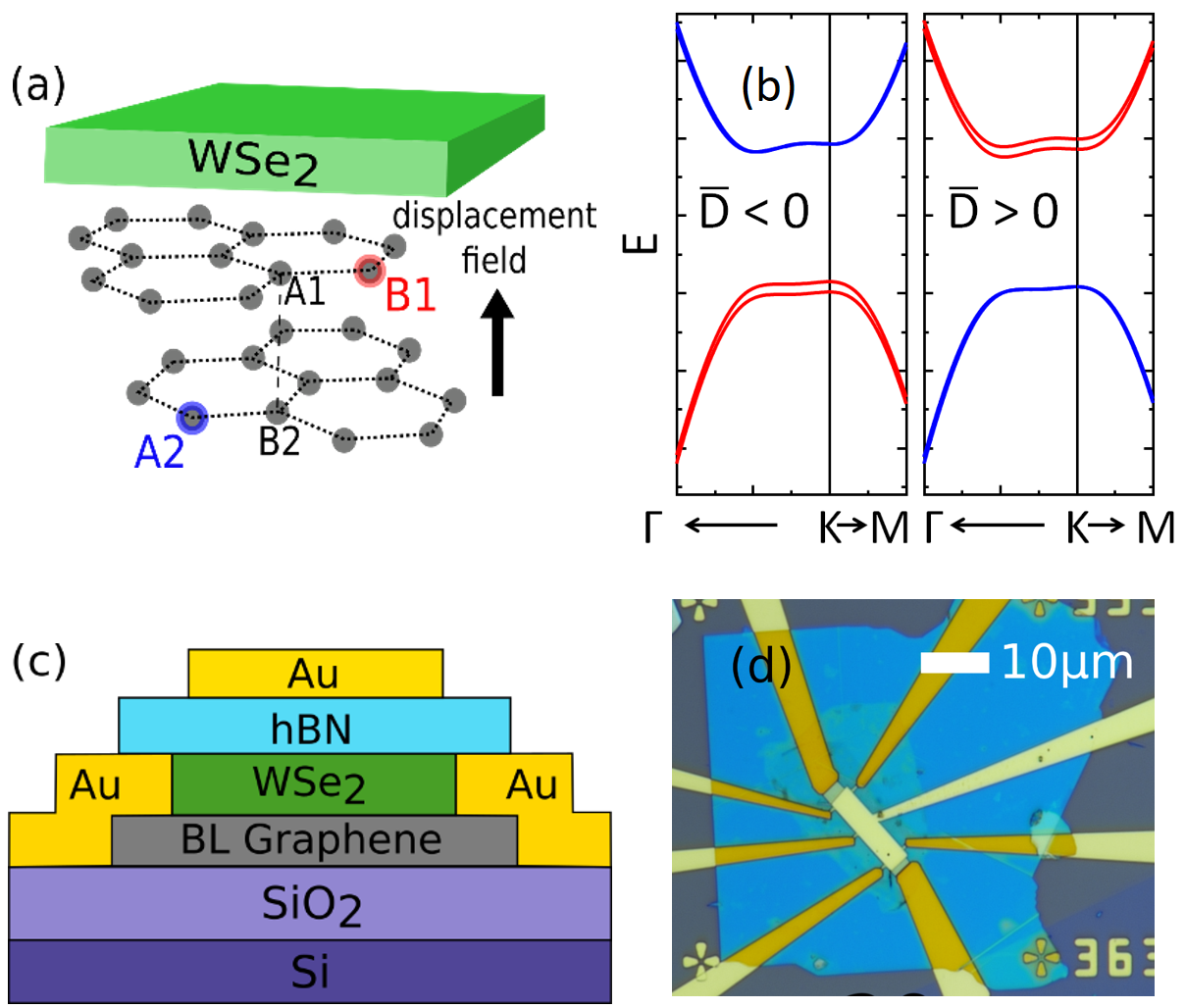}
	\caption[fabrication]{ (a) Schematic view of the two graphene layers under WSe$_2$. As layer 1 is closer to the TMD, the proximity induced SOC is expected to be much larger here than in layer 2. Additionally, the applied displacement field between the layers is shown, with the direction of the arrow for $\bar D>0$. Atoms B1 and A2 form the low-energy bands, the dimer atoms A1 and B2 are relevant for the full BLG Hamiltonian (see Appendix).	
		(b) Calculated band structure of BLG for opposite polarity of the displacement field, leading to spin splitting mainly in the valence band or conduction band. Predominant occupation of B1 and A2 atoms is indicated by red (B1) and blue (A2) lines. (c) Schematic cross-section of the devices.  (d) Optical microscope image of sample A. }
	\label{fig:fabrication}
\end{figure}

Spin-orbit coupling (SOC) is central to topological insulators \cite{Kane2004,Kane2005,Bernevig2006,Bernevig2006a,Konig2007} and spintronics, both very active fields of current research in condensed matter physics. To add flexibility in device functions, it is highly desirable to tune SOC by applying a gate voltage, as in the original proposal for a spin field effect transistor by Datta and Das \cite{DattaDas1990}. Graphene is an excellent material for spintronics, as its intrinsic SOC is very low \cite{Gmitra.2009}, leading to large spin lifetimes \cite{Han2014}. SOC can then be induced by proximity coupling to transition metal dichalcogenides \cite{Kaloni2014,Gmitra2016}, leading to dramatic modification of spin transport \cite{Cummings.2017,Ghiasi,Benitez,Garcia2017,Ghiasi2019,Safeer2019,Benitez2020}, or weak antilocalization (WAL) in magnetotransport \cite{Wang2015a,Wang2016,Yang2016,Voelkl2017,Zihlmann.2018,Wakamura2018,Afzal2018}. In addition, robust, spin-polarized edge channels were recently predicted for a graphene/WSe$_2$ heterostructure \cite{Frank2018a}. Here, again, gate control of SOC would be strongly beneficial. In single layer graphene (SLG)/TMDC heterostructures, gate tunability of SOC is only quite modest \cite{Gmitra.2015,Voelkl2017}. However, in a heterostructure of bilayer graphene (BLG) and a TMDC on one side only, strong gate tunability of SOC was recently predicted \cite{Gmitra.2017,Khoo.2017}. The underlying mechanism is quite intuitive: At low carrier densities, an out-of-plane displacement field $\bar D$ opens a gap in BLG, and leads to a selective population of one of the two layers only, with opposite trend for the valence band (VB) and conduction band (CB). More precisely, the occupancy of layers 1 and 2 is governed by 
the layer polarization $g_{1,2}$ \cite{Young2011,Khoo.2017}:
\begin{equation}
	g_{1,2}(\kappa,\bar D) = \dfrac{1}{2} \pm \dfrac{U(\bar D)}{\kappa \sqrt{4U(\bar D)^2 + \hbar^4 k_F^4 / m^{*2}}}.
	\label{Eq:Khoo}
\end{equation}
Here, $\pm U(\bar D)$ is the potential energy on layers 1 and 2, plus and minus sign correspond to layer 1 and 2, and $\kappa = \pm 1$ for valence and conduction band, respectively. 
Layer 1 is closer to WSe$_2$ (see Fig. \ref{fig:fabrication} (a)), thus the charge carriers in it experience a much stronger proximity-induced SOC than the carriers in the other layer.
Hence, the strength of induced SOC in BLG can be conveniently tuned by gate voltages. The behavior is opposite for electrons and holes due to the inversion of the bandstructure by the applied transverse electric field. 
The transition between high and low SOC strength is sharpest for low carrier densities, and softens considerably when going to higher densities.

 A full density functional calculation \cite{Gmitra.2017} of the band structure of BLG/TMDC shows that, depending on the field polarity, spin splitting is predominantly found in either VB or CB (see Fig. \ref{fig:fabrication} (b) for a corresponding tight-binding calculation). 
 The out-of-plane field $\bar D$ is composed of both built-in and gate-controlled fields. For details, we refer in particular to Fig. 4 of Ref. \cite{Gmitra.2017}.
 Recent charge transport and capacitance experiments confirmed the presence of an SOC gap, which depends on the gate voltage and carrier type \cite{Island2019,Wang2019}.

To confirm the implications of proximity-induced SOC on spin dynamics, WAL experiments are frequently employed \cite{McCann.2012b,Wang2015a,Wang2016,Yang2016,Voelkl2017,Zihlmann.2018,Wakamura2018,Afzal2018}. WAL is a phase-coherent effect, which is visible whenever the phase relaxation time $\tau_\varphi$ is long enough. In BLG, without SOC, weak localization (WL, visible as a dip in magnetoconductance around $B=0$) is found \cite{Bergmann1983,McCann.2006}, while the observation of WAL (a peak in magnetoconductance around $B=0$) confirms the presence of SOC, as WAL does not appear in BLG without strong SOC due to the Berry phase of $2\pi$ \cite{Kechedzhi2007}.
In the intuitive picture outlined above, one would therefore expect to observe a monotonic increase of the WAL feature when the electric field $\bar D$ is tuned such that spin splitting increases. In the following, we present a detailed experimental and theoretical study of WAL in BLG/WSe$_2$ showing that SOC is indeed strongly gate-tunable, as proposed recently, but the gate-dependence of WAL counterintuitively shows a non-monotonic dependence on $\bar D$. The experiments are confirmed by calculations of the spin relaxation rates using a model Hamiltonian of BLG with proximity-induced SOC on one layer only.

\section{Experiment}

In total, we present experimental data taken from three samples, A, B and C. Data from A and B are shown in the main text, additional data from the lower quality sample C are shown in the Supplemental Material \cite{SupplNote_Wse2BLG}.

\subsection{Sample fabrication}

Heterostructures were fabricated by using a hot pickup process with a polycarbonate film on polydimethylsiloxane (PDMS) \cite{Pizzocchero2016}.
A cross-section of the device is shown schematically in Fig. \ref{fig:fabrication}(c). Bernal stacked bilayer graphene was picked up by exfoliated WSe$_2$ with a thickness of a few tens of nanometers, and placed onto a standard p$^{++}$-doped Si/SiO$_2$ chip which functions as a back gate of the device. A Hall-bar structure was defined by electron-beam lithography (EBL) and reactive ion etching (RIE) with CHF$_3$/O$_2$ \cite{Wang2013}. Then, EBL and a RIE process with SF$_6$ were employed to selectively remove the WSe$_2$ at the contact areas \cite{Son2018}. The uncovered graphene was then contacted by 0.5 nm Cr / 40 nm Au. These contacts showed higher reliability and lower contact resistance compared to commonly used edge contacts. Hexagonal boron nitride (hBN) was placed on top of this device to serve as an additional dielectric for the subsequently deposited Au top gate. In Fig. \ref{fig:fabrication} (d) an optical microscope image of the finished sample A is depicted. With this dual gated device it is possible to independently tune charge carrier concentration and applied electric field \cite{Zhang2009,Oostinga2008}.

\begin{figure}
	\includegraphics[width=8.7cm]{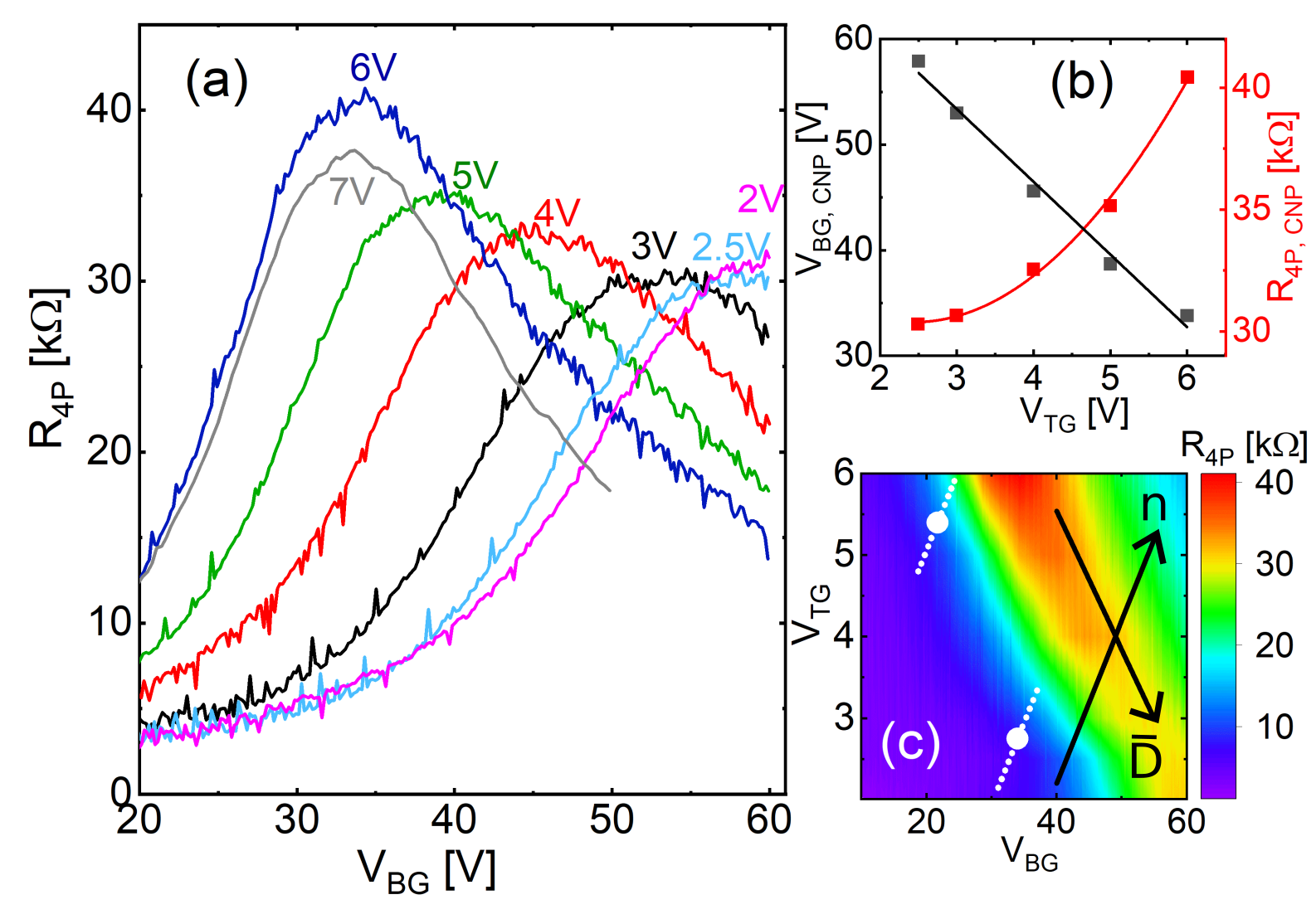}
	\caption[fabrication]{(a) Back gate sweeps at different top gate voltages
		. The CNP shifts with varying top gate values. (b) Dependence of the back gate voltage position and the resistance at the CNP on the top gate voltage. From these curves and the ratio of top  and back gate voltage at the CNP the gate characteristics can be calculated. (c) Gate map from the back gate sweeps in (a). The white dots show two measurement points around $n=1.3 \times 10^{16}$ m$^{-2}$ and two different displacement fields for the WAL measurements presented in Fig. \ref{fig:sigma13}. The dashed lines show the charge carrier density range for the 31 magnetoconductivity curves of which the average was taken. Also, the directions in which the charge carrier density and the displacement field can be increased are marked with arrows.}.
	\label{fig:char1}
\end{figure}

\subsection{Characterization}
All measurements were performed at a temperature of $T=1.4$ K with an AC current of $I_{AC}= 10\dots 50$ nA at a frequency of $f=13$ Hz. 
Figure \ref{fig:char1} (a) shows the four-point resistance $R_{4P}$ of sample A as a function of the back gate voltage for different top gate voltages. The corresponding gate map can be seen in Fig. \ref{fig:char1} (c). The back gate voltage position and maximum resistivity at the charge neutrality point (CNP) shift by varying the applied top gate voltage. This is caused by the electric field induced bandgap in bilayer graphene and the shift of the Fermi level via the applied top and back gate voltages \cite{Oostinga2008,Zhang2009}. The CNP position depends linearly on $V_{TG}$ shown by the black squares in Fig. \ref{fig:char1} (b). Further, the resistivity at the CNP shows a parabolic dependence on $V_{TG}$, shown in Fig. \ref{fig:char1} (b) with red squares. The minimum of the parabolic fit corresponds to the top gate/back gate combination ($V_{TG,0} = 2.5$ V, $V_{BG,0} = 57$ V), where the band gap in bilayer graphene vanishes, as both built-in electric fields and gate-controlled fields combine to a zero effective transverse electric field $\bar{D}$. 
From the combination of top and back gate capacitive coupling, we extract the required directions on the gate map
to change either carrier density $n$ or displacement field $\bar D$, shown as arrows in  Fig. \ref{fig:char1} (c). 
Here, $\bar D$ and $n$ are obtained in the following way \cite{Zhang2009}: 
\begin{equation}
	n= \dfrac{D_B - D_T}{e} \epsilon_0,\ \ 
	\bar{D} = \dfrac{D_B + D_T}{2},
\end{equation}
with
\begin{equation}
	D_B = \dfrac{V_{BG}-V_{BG,0}}{\epsilon_0} C_B,\ \ 
	D_T=-\dfrac{V_{TG}-V_{TG,0}}{\epsilon_0} C_T,
\end{equation}
where $C_B$ and $C_T$ are the capacitive couplings of back and top gate dielectrics, respectively. The polarity of $\bar D$ is positive for the electric field pointing from back gate to top gate, i.e., from BLG to TMDC.

The gate map depicted in Fig. \ref{fig:char1} (c) is limited by the back gate range, as an applied voltage higher than $ V_{BG}> \ $60 V might cause a breakdown of the back gate dielectric, destroying the sample. The CNP of the curve at a top gate voltage of 7 V in Fig. \ref{fig:char1} (a) does not follow the parabolic behavior of the other CNPs. The reason for this is that at top gate voltages higher than $V_{TG}>$ 6 V the Fermi level in WSe$_2$ is shifted into the conduction band and free electrons screen the top gate \cite{Gmitra.2017}. Therefore, the range of possible combinations of charge carrier densities and transverse electric fields was limited, and we could only study WAL in the hole regime of sample A.

The mobility on the hole-side of $\mu_h$ = 3000 \ cm$^2$/Vs can be extracted from the curve at 2.5 V top gate voltage in Fig. \ref{fig:char1}(a) which shows the lowest CNP. This leads to a mean free path of $l=40$ nm for a charge carrier density of $|n|=1.3 \times 10^{16}$ m$^{-2}$.
\begin{figure}
	\includegraphics[width=8.6cm]{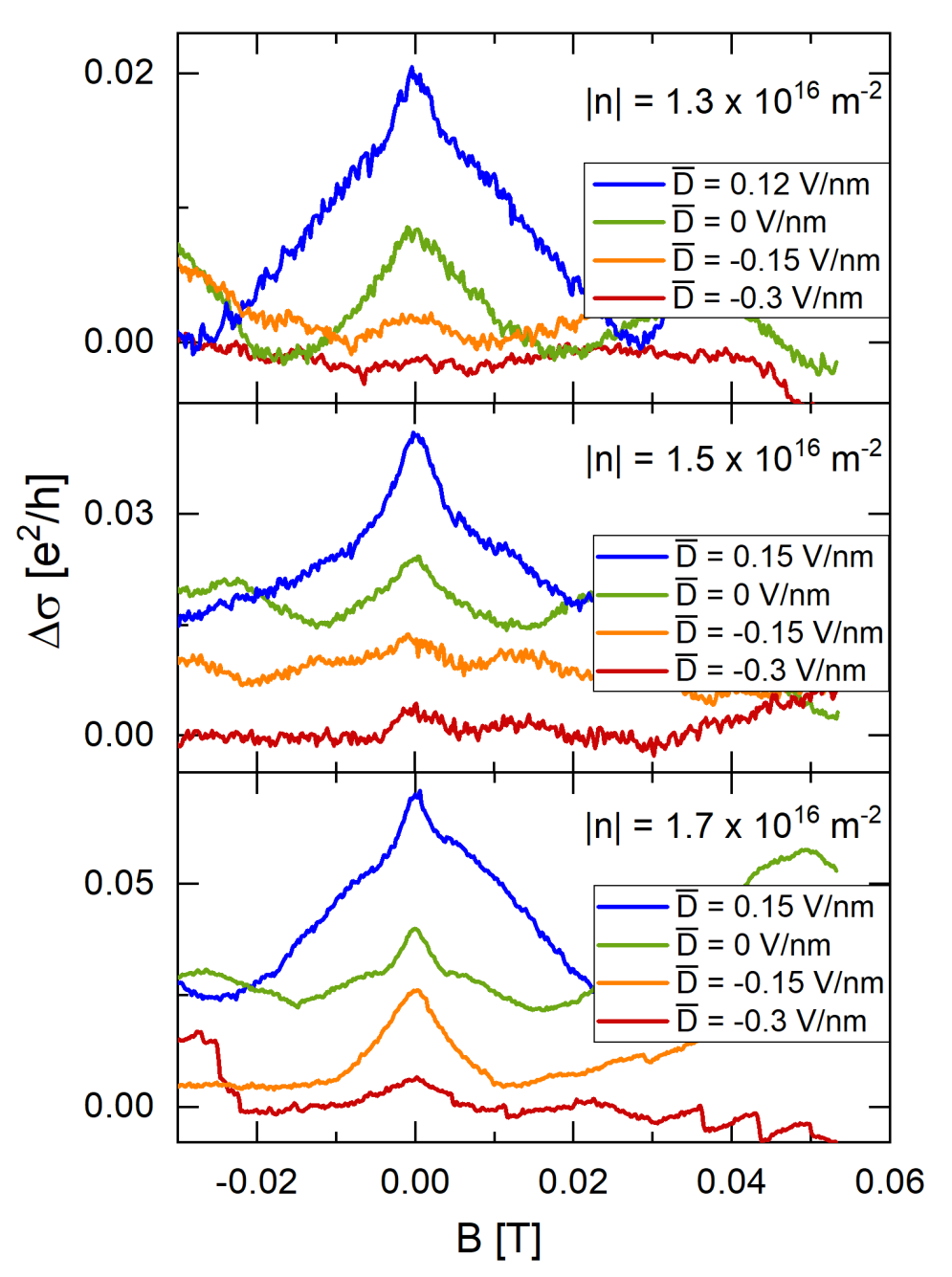}
	\caption[fabrication]{Gate-averaged magnetoconductivity of sample A at different transverse electric fields and average hole concentrations of $|n|=1.3,\ 1.5$, and $1.7 \times 10^{16}$ /m$^2$, taken at $T=1.4$ K. A clear dependence on the displacement field can be observed, as the peak height decreases with decreasing displacement fields. In the lowest panel, occasional jumps in the data are caused by slight gate instabilities.}
	\label{fig:sigma13}
\end{figure}

\subsection{Weak antilocalization of sample A}

The physics of WAL was calculated by McCann and Fal'ko for graphene subject to SOC terms in the Hamiltonian, which conserve or break $z\rightarrow -z$ symmetry \cite{McCann.2012b}. Those terms lead to the spin relaxation rates $\tau_{sym}^{-1}$ and $\tau_{asy}^{-1}$, respectively, which can, in principle, be determined by fitting the experimental data to the McCann theory. 
For a full description of WAL in SLG and BLG, the graphene-specific scattering times $\tau_{iv}$ (intervalley scattering time) and $\tau_*$ (which contains the intravalley scattering time $\tau_z$ by $\tau_*^{-1} = \tau_z^{-1} + \tau_{iv}^{-1}$) are included:
\begin{equation}
	\begin{split}
		\Delta \sigma (B)& = -\dfrac{e^2}{2\pi \hbar} \left[ F \left(  \dfrac{\tau_B^{-1}}{\tau_{\varphi}^{-1}}  \right) 
		- F \left(  \dfrac{\tau_B^{-1}}{\tau_{\varphi}^{-1} + 2\tau_{asy}^{-1}}  \right) \right.\\
		&- 2F \left(  \dfrac{\tau_B^{-1}}{\tau_{\varphi}^{-1} + \tau_{asy}^{-1}+ \tau_{sym}^{-1}}  \right) 
		- F \left(  \dfrac{\tau_B^{-1}}{\tau_{\varphi}^{-1} + 2\tau_{iv}^{-1}}  \right) \\
		& -2F \left(  \dfrac{\tau_B^{-1}}{\tau_{\varphi}^{-1} + \tau_{*}^{-1}}  \right)
		+F \left(  \dfrac{\tau_B^{-1}}{\tau_{\varphi}^{-1} + 2\tau_{iv}^{-1}+ 2\tau_{asy}^{-1}}  \right) \\
		&   +2F \left(  \dfrac{\tau_B^{-1}}{\tau_{\varphi}^{-1} + \tau_{*}^{-1} + 2\tau_{asy}^{-1}}  \right)  \\
		&+ 2F \left(  \dfrac{\tau_B^{-1}}{\tau_{\varphi}^{-1} + 2\tau_{iv}^{-1} + \tau_{asy}^{-1}+ \tau_{sym}^{-1}}  \right) \\		
		& \left. \pm 4F \left(  \dfrac{\tau_B^{-1}}{\tau_{\varphi}^{-1} + \tau_{*}^{-1}+\tau_{asy}^{-1}+ \tau_{sym}^{-1} }  \right)
		\right],
	\end{split}
	\label{Eq:conductivity}
\end{equation}
with $F(z) = \ln z +\psi\left(1/2 + 1/z\right)$, $\psi$ the digamma function, $\tau_B^{-1} = 4eDB_z/\hbar$ and the diffusion constant $D$ \cite{Zihlmann.2018,McCann.2012b,Iilic2019,McCann2021}. In the last row, the plus sign is for SLG, and the minus sign for BLG \cite{McCann2021}.

Inspection of this equation shows that $\tau_\varphi$ can be determined quite accurately, while the other scattering times are cross-dependent on one another in the fitting procedure. When WAL is observed (a peak in magnetoconductivity at $B=0$ T), SOC is certainly present. The absence of WAL, or presence of WL can be due to the lack of sizable SOC, or due to a long lifetime $\tau_{asy}$. The interpretation of experimental data is thus quite involved \cite{McCann.2012b,Zihlmann.2018}. Note that the above holds for BLG, where without SOC, WAL cannot be observed due to the Berry phase of $2\pi$ \cite{Kechedzhi2007}. In SLG, on the other hand, WAL can also be observed without SOC for certain combinations of elastic scattering times \cite{Tikhonenko2009}.

 Figure \ref{fig:sigma13} shows the dependence of the conductivity of sample A on an out-of plane magnetic field at different charge carrier concentrations and various displacement fields $\bar{D}$.
 As the scattering times $\tau_\varphi, \tau_{iv}$ and $\tau_{*}$ in Eq. \ref{Eq:conductivity} might depend on the carrier density, varying $\bar{D}$ at fixed carrier density enables us to single out the effect of an electric field on SOC. The magnetoconductivity curves are averaged over 31 measurements at slightly different charge carrier concentrations and constant transverse electric field to suppress universal conductance fluctuations (UCF) \cite{Gorbachev2007,Tikhonenko2008a,Wang2015a}.  A parabolic background fitted to reference measurements taken at $T=30$ K was subtracted. Here, for all curves a conductivity peak arises around $B=0$ T, stemming from WAL. 
 We observe a strong gate-dependence of WAL, confirming the gate-tunability of SOC, but the trend in $\bar D$ is opposite to what is expected: We find that in the hole regime, the WAL feature becomes more pronounced as $\bar D$ increases, but overall spin splitting should decrease in this direction. 
We note that fitting the experimental data to Eq. \ref{Eq:conductivity} is ambiguous, as widely varying values for $\tau_{sym}$ and $\tau_{asy}$ lead to acceptable fitting results (see Supplemental Material \cite{SupplNote_Wse2BLG} for ambiguity of fitting and also the suppression of WAL with increasing $T$ \cite{Narozhny2002}). For SLG/TMDC, it was proposed \cite{Zihlmann.2018} to fix the ratio of $\tau_{sym}$ and $\tau_{asy}$ to a value determined from spin precession experiments in spin valve devices \cite{Omar.2019}, but this approach is bound to fail in BLG/TMDC, as we will show below.

\section{Theory of the gate dependence of WAL}

To elucidate the counterintuitive gate dependence of WAL, we calculate the gate dependence of the spin relaxation times $\tau_{sym}$ and $\tau_{asy}$ using a model Hamiltonian describing BLG in one-sided contact to a TMDC layer. Those times are then inserted into Eq. \ref{Eq:conductivity}, together with the remaining scattering times extracted from experiment. 

\subsection{Calculation of spin relaxation times}

We calculate the spin relaxation following the procedure presented by Cummings {\em et al.}  \cite{Cummings.2017} for SLG/TMDC. Assuming the Dyakonov-Perel'-mechanism, symmetric and asymmetric rates are calculated by
\begin{equation}
	\tau_{sym}^{-1}=\left<\omega_z^2 \right>\tau_{iv},\ \ \tau_{asy}^{-1}=(\left<\omega_x^2 \right> + \left<\omega_y^2 \right>)\tau_{p},
	\label{Eq:rates}
\end{equation}
where $\tau_p$ is the momentum relaxation time, and the squares of the spin precession frequencies $\omega_{x,y,z}$ around the spin-orbit fields in $x$, $y$, and $z$ directions are averaged along the Fermi contour. Note that those rates do not directly correspond to the spin relaxation times for spins oriented along the $x$, $y$, or $z$-axis, as would be measured in a spin transport experiment \cite{Omar.2019,Note1}.
We obtain the precession frequencies $\omega_{x,y,z}$ for electrons and holes using the $8\times 8$ model Hamiltonian described in the Appendix. The strength of valley-Zeeman and Rashba SOC in the graphene layer close to WSe$_2$ is given by $\lambda_{VZ}$ and $\lambda_{R}$, respectively, while in the other graphene layer, SOC is set to zero. From the eight bands, we only consider the four innermost spin-resolved valence and conduction bands, which are composed predominantly from states residing on B1 and A2 (see Fig. \ref{fig:fabrication} (a)).
$U (\bar D)$ can be determined self-consistently from $\bar D$ \cite{McCann2013} (see Supplemental Material \cite{SupplNote_Wse2BLG}). Using an effective mass of $m^*=0.043$ times the free electron mass \cite{Li2016} and typical densities from our experiment, this results approximately in a relation $U=\bar D/10$ with the units in eV, and V/nm, respectively, which we use for simplicity in the following.
\begin{figure*}
	\includegraphics[width=16cm]{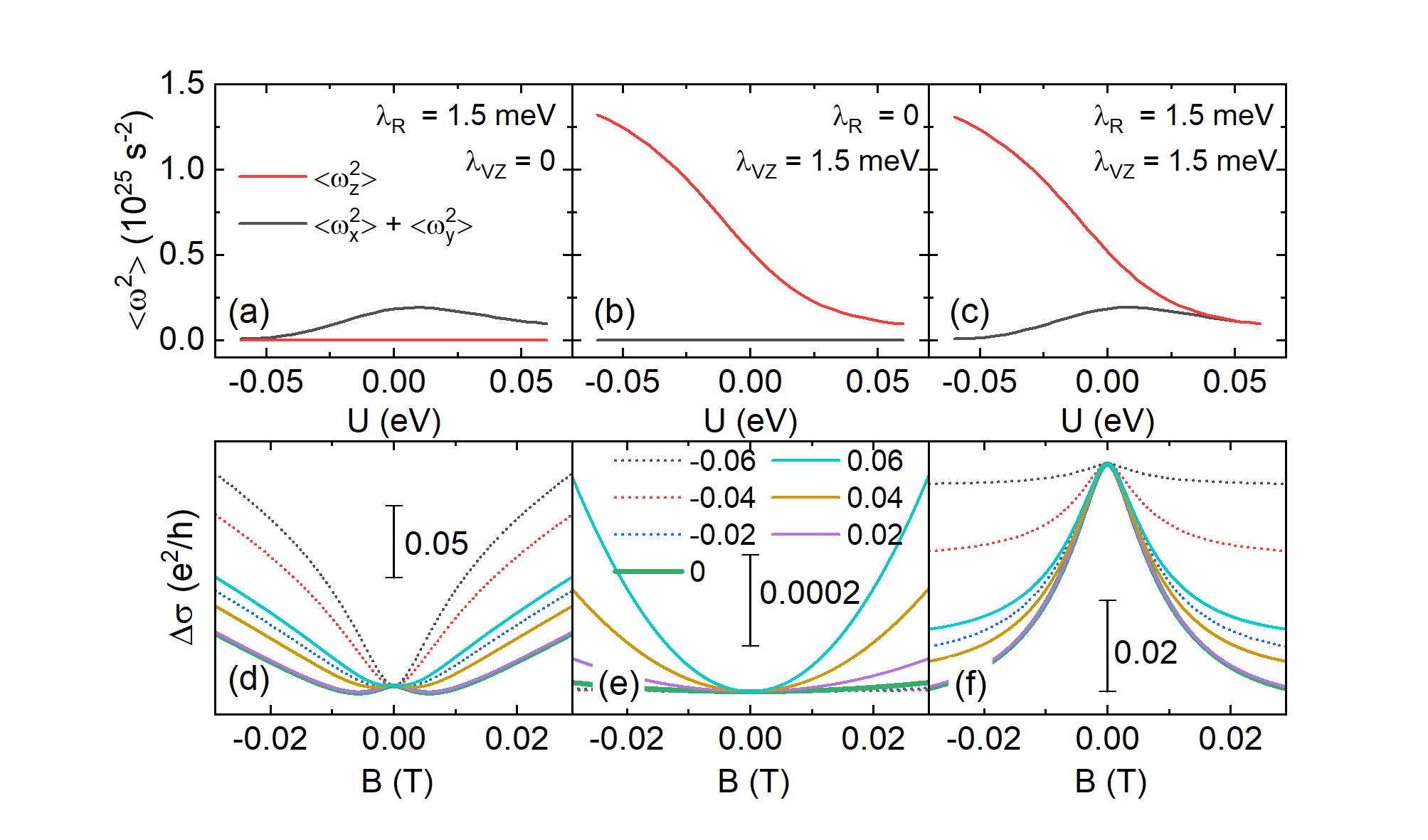}
	\caption{(a) to (c): Squared precession frequencies, averaged over the Fermi contour for the values of $\lambda_{VZ}$ and $\lambda_{R}$ as shown in the panels. (d) to (f) Corresponding WAL curves simulated for the layer potential $U$ in eV as shown in panel (e). The curve for $U=0$ eV is shown as a thicker line, for better visibility, and lines for $U<0$ eV are dotted.}
	\label{Fig:Prec_WAL}
\end{figure*}

The resulting gate dependence of the squared precession frequencies around $z$ and $x,y$ is plotted in Fig. \ref{Fig:Prec_WAL}(a) to (c) for several combinations of $\lambda_{VZ}$ and $\lambda_{R}$ and a carrier concentration of $|n|=1.5\times 10^{16}$ m$^{-2}$ of holes. We find that both quantities show a strong gate dependence, but with an opposite trend. While $\langle \omega_z^2\rangle$ decreases with $U$, thus following the expected trend, the precession around the $x$ and $y$ axes shows a maximum at $U=0$ eV. As the latter frequency enters into the asymmetric scattering rate $\tau_{asy}^{-1}$, which is essential to observing WAL, and not WL, this gives us a first clue to understanding the peculiar gate dependence of WAL. We also note that the ratio $\tau_{asy}/\tau_{sym}$ 
varies strongly with $U$ at a given density $n$. Furthermore, as expected, setting the valley-Zeeman term to zero suppresses precession around the $z$-axis (panel (a)), while setting the Rashba term to zero eliminates precession around the $x$ and $y$ axis (panel (b)). 

\subsection{Gate dependence of WAL}

The precession frequencies calculated in the previous section can now be translated into WAL curves. 
To simulate WAL in graphene, we take typical scattering times well suited to transport results from sample A: $\tau_{\varphi} = 7$ ps is extracted from the width of the WAL feature at $T=1.4$ K and is comparable to the phase coherence time obtained by taking the autocorrelation of UCFs (see Supplemental Material \cite{SupplNote_Wse2BLG} and also \cite{Pezzini2012,Lundeberg2012,Lundeberg2013}). $D=0.012$ m$^2$/s and $\tau_p = 72$ fs are obtained from typical mobilities of our samples, $\tau_{iv} \approx 14.5$ ps and $\tau_*  \approx 0.5$ ps are taken from reference samples where WSe$_2$ was replaced by hexagonal boron nitride on top of BLG. Note that $\tau_{iv}$ enters into $\tau_{sym}$ through Eq. \ref{Eq:rates}. The exact values of $\tau_{iv}$ and $\tau_*$ determine the curve shape, overall effect height, and bending of the curves at higher fields. Due to the large number of free parameters in Eq. \ref{Eq:conductivity}, these parameters cannot be determined unambiguously. However, the curves shown in Fig. \ref{Fig:Prec_WAL} reflect the trend of the gate dependence of WAL over a wide range of parameters. We observe that when only Rashba SOC is present (panel (d)), a transition from WL to WAL is found when increasing $U$ from negative values to $U=0$ eV, and the trend eventually reverses for $U>0$ eV. The transition from WL to WAL is still obtained when $\lambda_{VZ}\ne 0$, and $\lambda_{VZ}\ll \lambda_{R}$ (not shown). When $\lambda_{R}=0$ and only $\lambda_{VZ}\ne 0$, WAL is never found. The overall effect strength is tiny (note the scale bar in panel (e)) and WL strength can be slightly tuned with $U$.
When both $\lambda_{R}$ and $\lambda_{VZ}$ are non-zero, WAL is observed, and its strength can be tuned with $U$, but with a non-monotonic behavior. In our model, the highest WAL peak is observed for $U=0$, but this can be slightly modified when the graphene layer away from TMDC also bears a small SOC, as for intrinsic graphene. In the model presented in the Appendix, this small contribution is neglected. 

\begin{figure*}
	\includegraphics[width=16cm]{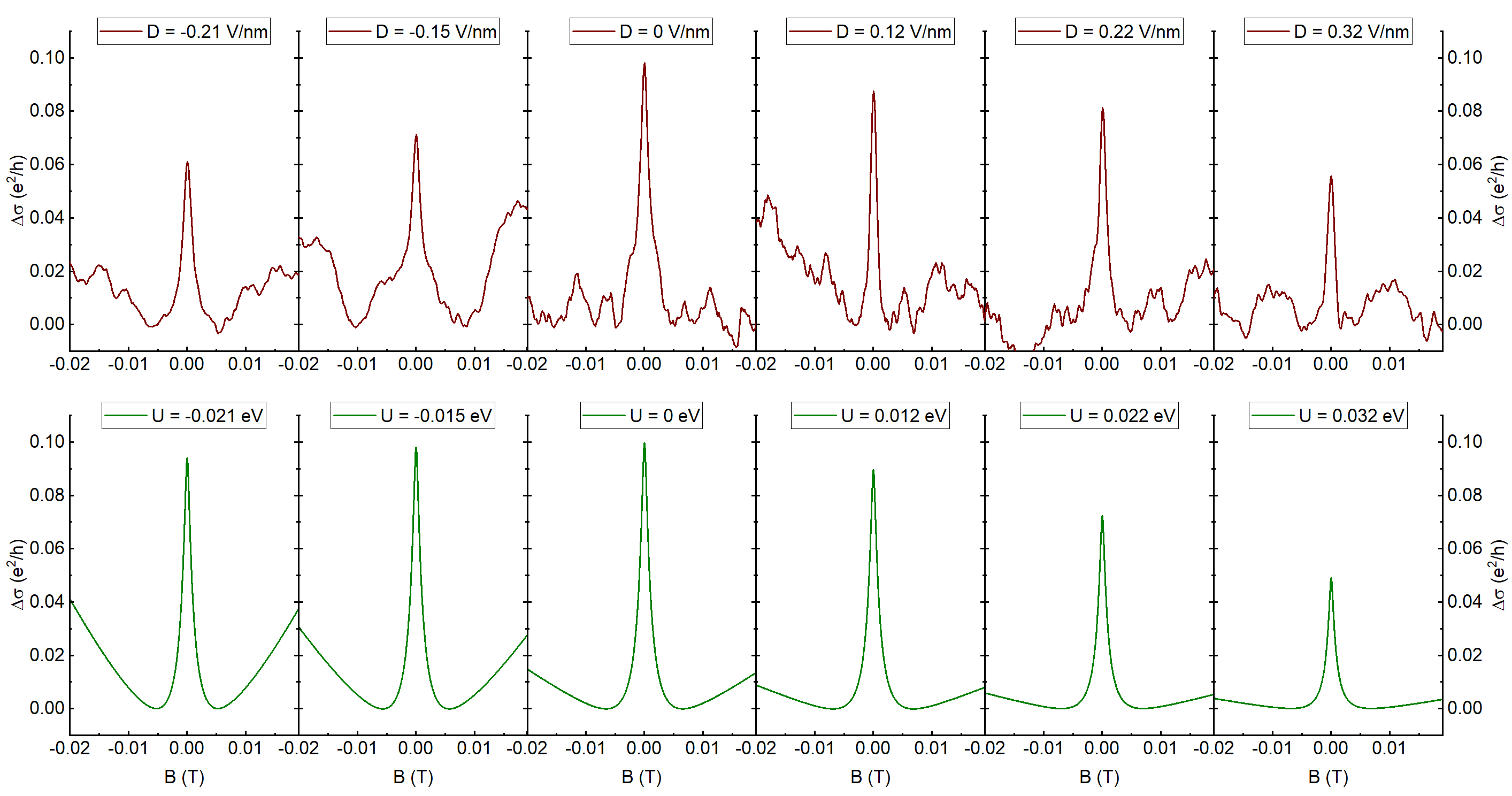}
	\caption{Comparison of experimental data for sample B, in the electron regime, at $n=1.3\times 10^{16}$ m$^{-2}$ at different out of plane electric fields $\bar D$ and simulations using scattering times and diffusion constant as determined from experiment, and $\lambda_{VZ}=0.8$ meV, $\lambda_{R}=0.5$ meV on the graphene layer close to WSe$_2$.}
	\label{Fig:ExptTheory}
\end{figure*}

Finally, we compare experimental data obtained from sample B, in the electron regime, with the results of our model (see Fig. \ref{Fig:ExptTheory}). As in sample B the charge neutrality point is found at smaller gate voltages, we can access both the electron and the hole regime. In the electron regime, $\bar D$ could be varied over a broader range so as to explore the behaviour on both sides of $\bar D=0$. For the modeling, we used $\tau_{iv} = 10$ ps, $\tau_p = 600$ fs, $\tau_z = 600$ fs, $\tau_\varphi= 12.1$ ps, $D = 0.090$ m$^2$/s, $n = 1.3 \times 10^{16}$ m$^{-2}$ (from transport characterization and reference samples) and $\lambda_{VZ} = 0.8$ meV, $\lambda_R = 0.5$ meV to obtain a visual match. The similarity is quite striking. Both experiment and theory show the most pronounced WAL feature at zero out of plane electric field with a drop in visibility on both sides \cite{Note2}.
This behavior is at odds with the simpler expectation gained from Eq. \ref{Eq:Khoo}, but is the result of the intricate dependence of Eq. \ref{Eq:conductivity} on two different, SOC-based relaxation rates. Mainly, as WAL in graphene with SOC can only be observed if a sizable rate $\tau_{asy}^{-1}$ is present, its non-monotonic dependence on the out-of-plane electric field directly determines the non-monotonic evolution of WAL with the gate voltage. The values for the SOC strength $\lambda_{VZ}$ and $\lambda_R$ are in good agreement with results from density functional theory. Moreover, using the same $\lambda_{VZ}$ and $\lambda_R$, we can also achieve good agreement for WAL data in the hole regime, and the electron regime at a different density (see Supplemental Material \cite{SupplNote_Wse2BLG}).

\section{Discussion and Conclusion}

Our results show that, while SOC is indeed strongly gate tunable in BLG/TMDC heterostructures, the interpretation of WAL experiments is not straightforward. Unlike the SLG/TMDC case, or the TMDC/BLG/TMDC heterostructure (which we both checked in our formalism using appropriate model Hamiltonians), the spin splittings projected to the $x$,$y$ and $z$-directions are strongly density and gate dependent. In particular, the average $\left<\omega_x^2 \right> + \left<\omega_y^2 \right>$ along the Fermi contour is non-monotonic in the out-of-plane electric field, and enters into the scattering rate $\tau_{asy}^{-1}$. The latter is crucial for observing WAL in graphene with SOC. When only $\tau_{sym}^{-1}$ is relevant, one still observes WL with a reduced amplitude, even though SOC is present \cite{McCann.2012b}. In the simplest case, when only proximity induced SOC in one graphene layer is considered, the maximum visibility of WAL is present at precisely $U=0$ (corresponding to $\bar D = 0$ in experiment). On the other hand, setting SOC in the other graphene layer to small values, as in the pristine graphene case, can shift this maximum slightly away from $U=0$, but the general trend remains. Lastly, we briefly comment on a recent experimental demonstration of gate-tunable SOC in a similiar heterostructure by Tiwari {\em et al.} \cite{Tiwari2021}. Their results are not in line with the experimental and theoretical data in this study. We note, however, that in their article there is an inconsistency regarding the definition of the field direction between the formula in the text and the corresponding illustration.

In conclusion, we demonstrate a clear gate tunability of WAL in bilayer graphene, in proximity to WSe$_2$ on one side, which can be fully explained by calculations of the SOC induced spin relaxation. The fact that the visibility of WAL does not follow the gate-induced electric field monotonically, as might be expected at first sight, is accounted for by the intricate dependence of WAL in graphene on two SOC related scattering rates, and their non-monotonic dependence on the gate voltages.

\begin{acknowledgments}
	K.W. and T.T. acknowledge support from the Elemental Strategy Initiative
	conducted by the MEXT, Japan, Grant Number JPMXP0112101001,  JSPS
	KAKENHI Grant Number JP20H00354 and the CREST(JPMJCR15F3), JST.
	This work was funded by the Deutsche Forschungsgemeinschaft (DFG, German Research Foundation) through SFB 1277 (project-id 314695032, subprojects A09, B07), GRK 1570, SFB 689 and ER 612/2 (project id. 426094608).
	We thank T. Frank for fruitful discussions. 
	We are particularly  grateful to C.  Stampfer and E. Icking for enlightening discussions of the sign of $\bar D$. 
\end{acknowledgments}

\appendix
\section{Hamiltonian}

The electronic band structure of bilayer graphene in proximity to WSe$_2$ is obtained from the full $8\times 8$ Hamiltonian in the spirit of Refs.~\cite{McCann2013,Konschuh2012}. Its orbital part counts the relevant nearest intra- and inter-layer hoppings called conventionally as $\gamma_0=2.6$\,eV, $\gamma_1=0.34$\,eV, $\gamma_3=0.28$\,eV and $\gamma_4=-0.14$\,eV including also the dimer energy-shift $\Delta=9.7$\,meV. The above values are taken from Ref.~\cite{Konschuh2012}. The effect of the transverse displacement field on the electronic states in the 
top and bottom layers is captured by the on-site potential energy $U$. In the experiment the built-in dipole field of the bilayer due to the WSe$_2$ is counteracted by the external gates and from that reference configuration one measures displacement field $\bar{D}$ and hence 
also the potential energy $U$. For those reasons the model Hamiltonian does not involve any dipole fields. 
The full orbital Hamiltonian $H_{o}$ in the Bloch basis based on atomic orbitals $A_{2\uparrow}$, $A_{2\downarrow}$, $B_{1\uparrow}$, $B_{1\downarrow}$, $A_{1\uparrow}$, $A_{1\downarrow}$, $B_{2\uparrow}$, $B_{2\downarrow}$, for their atomic positions see Fig.~1(a), 
is given as:
\begin{widetext}
	\begin{equation}
		H_{o}(\mathbf{k})=\left(
		\begin{array}{cccccccc}
			-U & 0 & e^{i\mathbf{k}\cdot\mathbf{a}_3} f(\mathbf{k})\gamma_3 & 0 
			& \overline{f(\mathbf{k})}\gamma_4 & 0 & -\overline{f(\mathbf{k})}\gamma_0 & 0\\
			0 & -U & 0 & e^{i\mathbf{k}\cdot\mathbf{a}_3} f(\mathbf{k})\gamma_3 & 0 
			& \overline{f(\mathbf{k})}\gamma_4 & 0 & -\overline{f(\mathbf{k})}\gamma_0 \\
			e^{-i\mathbf{k}\cdot\mathbf{a}_3} \overline{f(\mathbf{k})}\gamma_3 & 0 & U & 0
			& -f(\mathbf{k})\gamma_0 & 0& f(\mathbf{k})\gamma_4 &  0 \\
			0 & e^{-i\mathbf{k}\cdot\mathbf{a}_3} \overline{f(\mathbf{k})}\gamma_3 & 0 & U 
			& 0 & -f(\mathbf{k})\gamma_0 & 0 & f(\mathbf{k})\gamma_4\\
			f(\mathbf{k})\gamma_4 & 0 & -\overline{f(\mathbf{k})}\gamma_0 & 0
			&  U +\Delta & 0 & \gamma_1 & 0\\
			0 & f(\mathbf{k})\gamma_4 & 0 & -\overline{f(\mathbf{k})}\gamma_0 & 0
			&  U +\Delta  & 0 & \gamma_1 \\
			-f(\mathbf{k})\gamma_0 & 0 & \overline{f(\mathbf{k})}\gamma_4 & 0 
			& \gamma_1 & 0 & -U+\Delta & 0\\
			0 & -f(\mathbf{k})\gamma_0 & 0 & \overline{f(\mathbf{k})}\gamma_4 & 0 
			& \gamma_1 & 0 & -U+\Delta
		\end{array} 
		\right).
	\end{equation}
\end{widetext}
In the above expression $\mathbf{k}$ is measured from the $\Gamma$ point, and
\begin{equation}
	f(\mathbf{k})=1+e^{i\mathbf{k}\cdot\mathbf{a}_2}+e^{-i\mathbf{k}\cdot\mathbf{a}_3},
\end{equation}
is the tight-binding structural function including the lattice vectors $\mathbf{a}_{j=1,2,3}=a\left(\cos{\tfrac{2\pi}{3}(j-1)},\sin{\tfrac{2\pi}{3}(j-1)}\right)$ and the graphene
lattice constant $a=0.246$\,nm.

To model the spin-orbit part $H_s$ we assume only the spin-orbit couplings in the proximitized top layer, namely, the next-nearest neighbor spin conserving VZ-term $\lambda_{VZ}$, and the nearest neighbor spin flipping Rashba coupling $\lambda_R$. These are assumed to be enhanced due to the proximity of WSe$_2$, and close to the Dirac points $\mathbf{K}^{\pm}=\pm\frac{4\pi}{3a}(1,0)$ they are primarily responsible for splitting of the electronic bands. 
The corresponding $H_s$ given in the same basis $A_{2\uparrow}$, $A_{2\downarrow}$, $B_{1\uparrow}$, $B_{1\downarrow}$, $A_{1\uparrow}$, $A_{1\downarrow}$, $B_{2\uparrow}$, $B_{2\downarrow}$ is as follows:
\begin{widetext}
	\begin{equation}
		H_s(\mathbf{k})=\left(
		\begin{array}{cccccccc}
			0 & 0 & 0 & 0 & 0 & 0 & 0 & 0\\
			0 & 0 & 0 & 0 & 0 & 0 & 0 & 0\\
			0 & 0 & f_{I}(\mathbf{k})\lambda_{VZ} & 0 & 0 & i f_{1R}(\mathbf{k})\lambda_R & 0 &  0 \\
			0 & 0 & 0 & - f_{I}(\mathbf{k})\lambda_{VZ} & i \overline{f_{2R}(\mathbf{k})}\lambda_R & 0 & 0 & 0 \\
			0 & 0 & 0 & -i f_{2R}(\mathbf{k})\lambda_R &  f_{I}(\mathbf{k})\lambda_{VZ} & 0 & 0 & 0\\
			0 & 0 &  -i \overline{f_{1R}(\mathbf{k})}\lambda_R & 0 & 0
			& - f_{I}(\mathbf{k})\lambda_{VZ} & 0 & 0 \\
			0 & 0 & 0 & 0 & 0 & 0 & 0 & 0\\
			0 & 0 & 0 & 0 & 0 & 0 & 0 & 0\\
		\end{array} 
		\right),
	\end{equation}
and the corresponding tight-binding functions read:
\begin{align}
	f_{I}(\mathbf{k})&=\frac{\sin{(\mathbf{k}\cdot\mathbf{a}_1})+\sin{(\mathbf{k}\cdot\mathbf{a}_2})+
		\sin{(\mathbf{k}\cdot\mathbf{a}_3})}{\frac{3}{2}\sqrt{3}},\\
	f_{1R}(\mathbf{k})&=\frac{2}{3}\left(1+e^{i(\frac{2\pi}{3}+\mathbf{k}\cdot\mathbf{a}_2)}+e^{-i(\frac{2\pi}{3}+\mathbf{k}\cdot\mathbf{a}_3)}\right),\\
	f_{2R}(\mathbf{k})&=\frac{2}{3}\left(1+e^{i(\frac{2\pi}{3}-\mathbf{k}\cdot\mathbf{a}_2)}+e^{-i(\frac{2\pi}{3}-\mathbf{k}\cdot\mathbf{a}_3)}\right).
\end{align}
The model Hamiltonian $H_o(\mathbf{k})+H_s(\mathbf{k})$ with the parameters $\lambda_{VZ}$ and $\lambda_{R}$ is used to calculate the spin-resolved bands and the spin relaxation rates.
The layer dependent energy term $U$ can be determined self-consistently from $\bar D$ \cite{McCann2013} (see Supplemental Material \cite{SupplNote_Wse2BLG}). Using an effective mass of $m^*=0.043$ times the free electron mass \cite{Li2016}, for simplicity we approximate $U(\bar D)$ by  a relation $U=\bar D/10$ with the units in eV, and V/nm, respectively.
\end{widetext}


%



\section*{Supplementary material}

# Supplemental Material: Counterintuitive gate dependence of weak antilocalization in bilayer graphene/WSe$_2$ heterostructures

Julia Amann,[1] Tobias Völkl,[1, *] Tobias Rockinger,[1] Denis Kochan,[2] Kenji Watanabe,[3] Takashi Taniguchi,[4] Jaroslav Fabian,[2] Dieter Weiss,[1] and Jonathan Eroms[1, †]

[1]*Institut für Experimentelle und Angewandte Physik,*
*Universität Regensburg, 93040 Regensburg, Germany*
[2]*Institut für Theoretische Physik, Universität Regensburg, 93040 Regensburg, Germany*
[3]*Research Center for Functional Materials, National Institute for Materials Science, 1-1 Namiki, Tsukuba 305-0044, Japan*
[4]*International Center for Materials Nanoarchitectonics,*
*National Institute for Materials Science, 1-1 Namiki, Tsukuba 305-0044, Japan*
(Dated: February 22, 2022)

## I. SELF-CONSISTENT CALCULATION OF THE BAND GAP

To relate the displacement field $\bar{D}$, applied by external gate voltages, and the potential difference $U' = 2U$ between the two graphene layers, we follow Eq. (74) in the review paper by McCann and Koshino, [1] which we reproduce here for convenience:

$$\frac{U'(n)}{U'_{ext}} \approx \left[1 - \frac{\Lambda}{2} \ln\left(\frac{|n|}{2n_\perp} + \frac{1}{2}\sqrt{\left(\frac{n}{n_\perp}\right)^2 + \left(\frac{U'}{2\gamma_1}\right)^2}\right)\right]^{-1}, \quad (1)$$

with

$$n_\perp = \frac{\gamma_1^2}{\pi \hbar^2 v^2}, \qquad \Lambda = \frac{c_0 e^2 n_\perp}{2\gamma_1 \epsilon_0 \epsilon_r}. \quad (2)$$

For the interlayer hopping parameter $\gamma_1$ we use 0.34 eV, $\epsilon_r \approx 1$, and $v \approx 10^6$ m/s corresponds to the Fermi velocity of single layer graphene. The expression is used to calculate the density potential difference $U'(n)$, including screening, self-consistently, when an external potential difference

$$U'_{ext} = e c_0 \bar{D} \quad (3)$$

is applied through gate voltages. $c_0 = 3.35$ Å is the bilayer graphene interlayer spacing. The background charges are included in the offset of back and top gate voltage, $V_{BG,0}$ and $V_{TG,0}$. The equation is solved self-consistently by iteration. As we do not strive to obtain a full, quantitative match of the calculations to the experiment, we content ourselves with a simplified estimate of $U = \bar{D}/10$, when the units are eV, and V/nm, respectively

## II. TEMPERATURE DEPENDENCE OF SAMPLE A

To check if the effect we are investigating is phase coherent, we compare averaged magnetoconductivities at a fixed measurement point at $|n| = 1.5 \times 10^{16}$ m$^{-2}$ and a transverse field of $\bar{D} = $ -0.15 V/nm for different temperatures in Fig. S1. As expected [2], due to reduced phase coherence, the WAL peak height decreases with increasing temperature. Due to the ambiguity of fitting (see below), we do not include a fit to the expression for WAL in this graph.

---

* Present address: Department of Condensed Matter Physics, Weizmann Institute of Science, Rehovot, Israel
† jonathan.eroms@ur.de

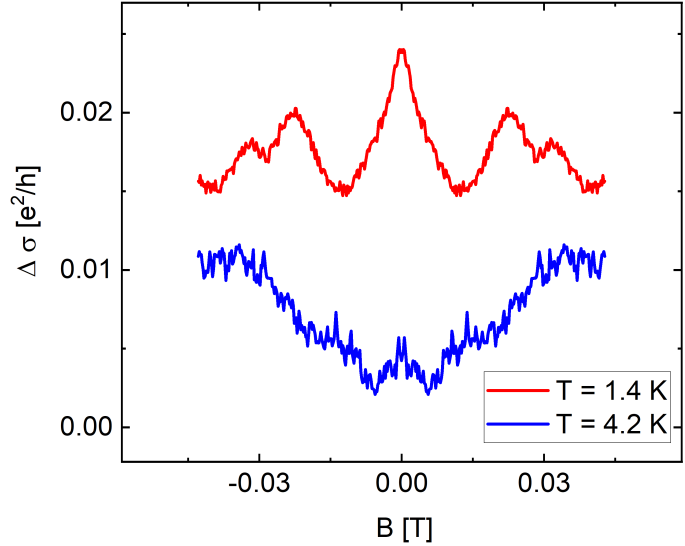

FIG. S 1: Gate-averaged magnetoconductivity of sample A at $|n| = 1.5 \times 10^{16}$ m$^{-2}$ and a transverse field of $\bar{D} =$ -0.15 V/nm at different temperatures. The WAL peak height is decreasing with increasing temperatures.

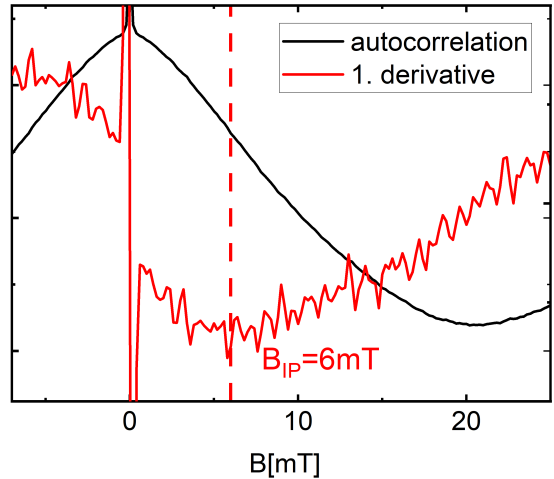

FIG. S 2: Autocorrelation of universal conduction fluctuations in sample A. The autocorrelation from 31 magnetoconductivity curves at $n = 1.3 \times 10^{16}$ m$^{-2}$ and $\bar{D} = 0.15$ V/nm is shown in black. The red curve is its first derivative, both are without scale. From the inflection point at 6 mT we can estimate the phase coherence time for the sample.

### III. PHASE COHERENCE TIME FROM THE AUTOCORRELATION FUNCTION OF SAMPLE A

To estimate the phase coherence time one can use the autocorrelation function. For this, the autocorrelation was taken for 31 magnetoconductivity curves consecutively for the same displacement field $\bar{D} = 0.15$ V/nm and slightly varying charge carrier density around $|n| = 1.3 \times 10^{16}$ m$^{-2}$. In Fig. S2, the function is shown in black with its derivative in red. From the inflection point at 6 mT we can calculate the estimated phase coherence time: [3–5]:

$$\tau_{UCF} = \tau_\varphi \approx \frac{3\hbar}{2eDB_{IP}} = \frac{3\hbar}{2e \cdot 0.0122 \frac{\text{m}^2}{\text{s}} \cdot 0.006 \text{ T}} = 13.5 \text{ ps}. \qquad (4)$$

This phase coherence time is on the same order of magnitude as the times extracted from WAL.

## IV. AMBIGUITY OF DIRECT FITTING, SAMPLE A

In Fig. S3 the averaged conductivity at $|n| = 1.5 \times 10^{16}$ m$^{-2}$ and a transverse field of $\bar{D} = 0$ V/nm at 1.4 K is shown. The green, red and blue curves are fits that were made at fixed $\tau_{iv} = 14.5$ ps and $\tau_*$ exlcuded, with a range of spin relaxation and phase coherence times, which can be seen in the table on the right. Here, the anisotropy $\eta = \tau_{asy}/\tau_{sym}$ varies over several orders of magnitude, but all fits seem to describe the magnetoconductivity well. We conclude that fitting the WAL data to theory does not allow us to extract the SOC related scattering times with certainty.

## V. TRANSPORT PARAMETERS

For sample A, we determined a hole mobility of $\mu_{el} = 0.3$ m$^2$/Vs ( = 3000 cm$^2$/Vs) from the gate dependence of the conductivity at $\bar{D} = 0$. At a density of $|n| = 1.5 \times 10^{16}$ m$^{-2}$, this results in a mean free path of $l_{mfp} = 42$ nm, a Fermi velocity of $v_F = 580\,000$ m/s (using $m_* = 0.043\, m_0$) and a diffusion constant of $D = \frac{1}{2} v_F l_{mfp} = 0.0122$ m$^2$/s. From WL measurements in a diffusive reference sample without WSe$_2$ we obtained $L_{iv} = 420$ nm, and thus $\tau_{iv} = 14.5$ ps. $\tau_*$ in the WAL formula is of little importance to the overall gate response of WAL. From reference measurements we found it to be in or below the ps range. We set it to 0.5 ps, and $\tau_\varphi$ was set to 7 ps for the calculations for sample A (Fig. 4 of the main text). The latter is in the same range as $\tau_\varphi$ from universal conductance fluctuations (see above).

For sample B, we determined $\mu_h = 1.9$ m$^2$/Vs ( = 19 000 cm$^2$/Vs) in the hole, and $\mu_{el} = 2.5$ m$^2$/Vs ( = 25 000 cm$^2$/Vs) in the electron regime. Using the respective densities for the data in the main text and the Supplemental Material, we obtain ($n < 0$ for holes):

| $n$ | $l_{mfp}$ | $v_F$ | $D$ | $\tau_p$ | $\tau_\varphi$ |
|---|---|---|---|---|---|
| $1.0 \times 10^{16}$ m$^{-2}$ | 290 nm | 477 000 m/s | 0.069 m$^2$/s | 600 fs | 11.7 ps |
| $1.3 \times 10^{16}$ m$^{-2}$ | 330 nm | 544 000 m/s | 0.090 m$^2$/s | 600 fs | 12.1 ps |
| $-1.6 \times 10^{16}$ m$^{-2}$ | 280 nm | 600 000 m/s | 0.084 m$^2$/s | 470 fs | 8.3 ps |

The phase coherence time was determined by matching the width of the highest WAL peak in each series, and corresponds reasonably to $\tau_\varphi$ determined from the autocorrelation of UCFs of sample B (22 ps). The intervalley scattering time $\tau_{iv}$ enters both into the visibility of WAL and in the symmetric spin lifetimes $\tau_{sym}$. Taking the order of magnitude from reference measurements, we fixed it to $\tau_{iv} = 10$ ps in all fits. Finally, it is reasonable to assume that the intravalley scattering time $\tau_z$ is not lower than the momentum relaxation time $\tau_p$. For sample B we set $\tau_z = 600$ fs in all fits. $\tau_*^{-1} = \tau_z^{-1} + \tau_{iv}^{-1}$ then enters into Eq. (4) of the main text.

## VI. ADDITIONAL DATA FROM SAMPLE B

In addition to the experimental curves shown in the main text, we provide further measurements at a different electron density of $n = 1.0 \times 10^{16}$ m$^{-2}$ (see Fig. S4) and in the hole regime, at $n = -1.6 \times 10^{16}$ m$^{-2}$ (see Fig. S5). Corresponding calculations of the WAL curves were performed with identical $\lambda_{VZ}$ and $\lambda_R$, and further scattering times as in the table above. Altogether, the good match between experiment and calculations in Fig. 5 of the main text and Figs. S4 and S5 further underscores the validity of our theory.

## VII. COMPARISON OF TUNABILITY IN ELECTRONS AND HOLES IN SAMPLE C

In Fig. S6, the magnetoconductivity of sample C is shown at a charge carrier density of $|n| = 1 \times 10^{16}$ m$^{-2}$ for both electrons and holes. Those curves are also averaged over 31 measurements at slightly varying charge carrier densities and constant transverse electric field. We subtracted a parabolic background fitted to reference data taken at $T = 10$ K. The black curves in Fig. S6 are shown to clarify the WAL peaks around $B = 0$ T and do not lead to unambiguous scattering times in the fit formula Eq. (4) from the main text. They should be considered as guides to the eye. This is due to possible superposition with UCFs, so that the WAL peak size is not clearly identifiable. Therefore it was not possible to calculate the SOC-strengths for this sample.

Nevertheless the response of the WAL peak to an electric field is opposite for electrons and holes: for electrons a

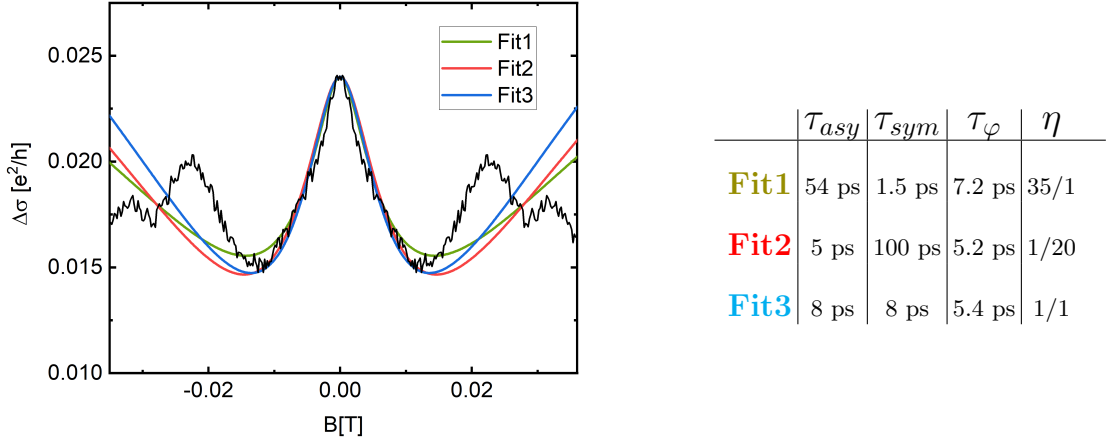

FIG. S 3: Gate-averaged magnetoconductivity at $n = 1.5 \times 10^{16}$ m$^{-2}$ and a transverse field of $\bar{D} = 0$ V/nm at 1.4 K. The WAL peak was fitted with a range of parameters that are shown in the table on the right, regardless of the anisotropy $\eta = \tau_{asy}/\tau_{sym}$. As all of the three fits in green, red and blue seem to describe the magnetoconductivity well, it is impossible to choose the correct combination.

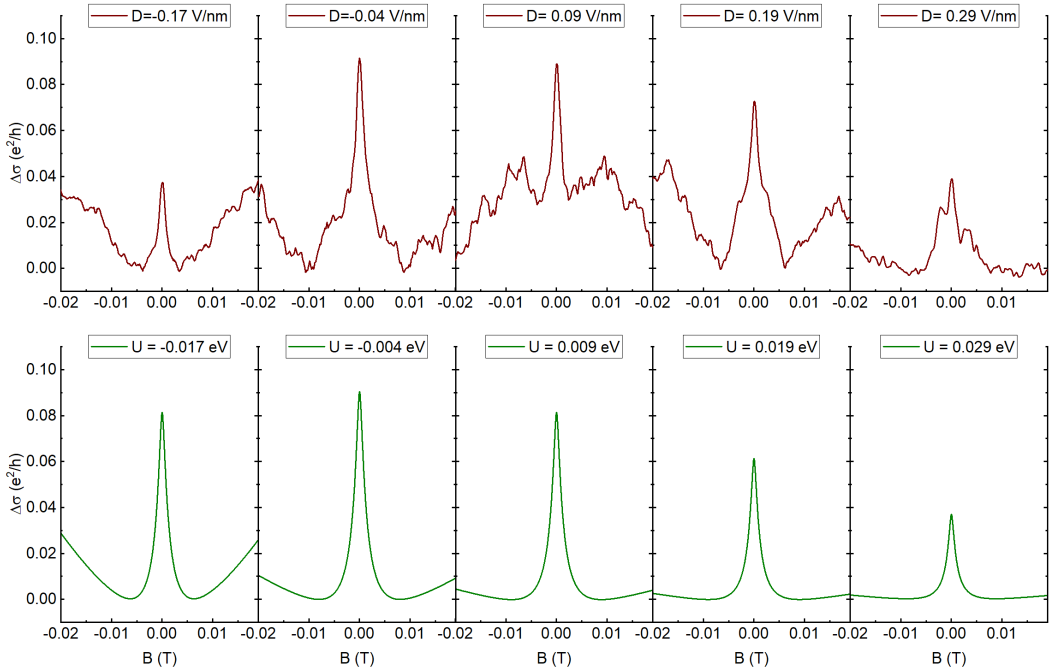

FIG. S 4: Sample B, comparison of experimental WAL curves and calculated ones at an electron density of $n = 1.0 \times 10^{16}$ m$^{-2}$. The parameters for $\lambda_{VZ} = 0.8$ meV and $\lambda_R = 0.5$ meV are the same as in Fig. 5 of the main text.

peak is visible at a negative displacement field of $\bar{D} = -0.15$ V/nm, whereas at the hole side peaks appear at positive displacement fields of $\bar{D} = 0$ V/nm and $\bar{D} = 0.15$ V/nm. Again, this behavior does not match the simple expectation from the layer polarization picture.

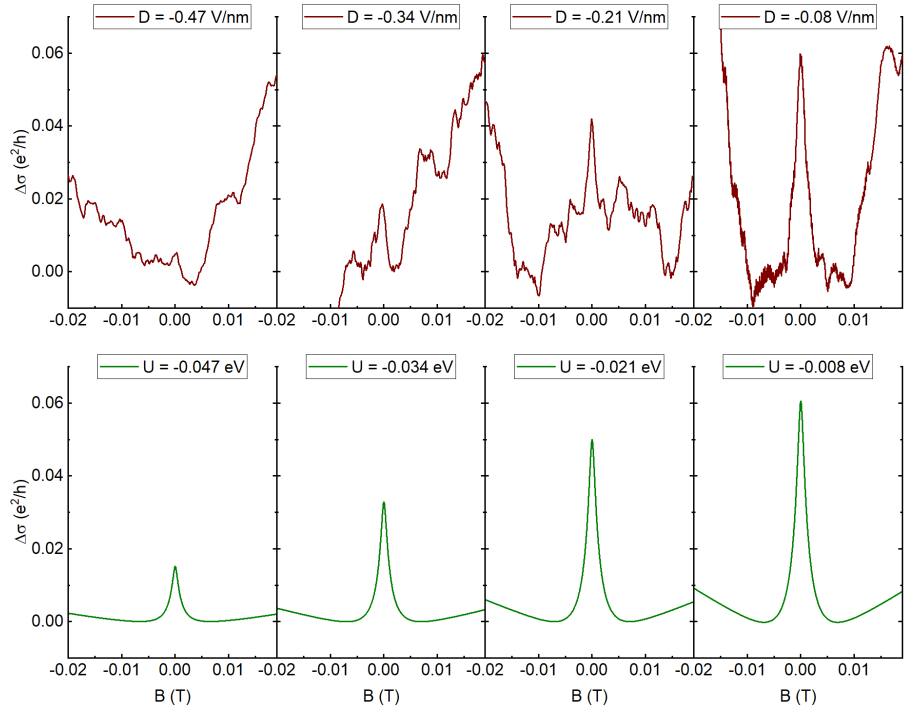

FIG. S 5: Sample B, comparison of experimental WAL curves and calculated ones at a hole density of $|n| = 1.6 \times 10^{16}$ m$^{-2}$. The parameters for $\lambda_{VZ} = 0.8$ meV and $\lambda_R = 0.5$ meV are the same as in Fig. 5 of the main text. The general trend and approximate peak heights match well.

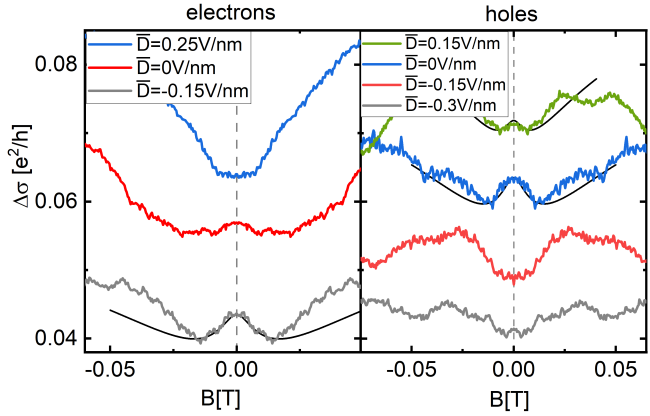

FIG. S 6: Gate-averaged magnetoconductivity for sample C, at different transverse electric fields and a charge carrier concentration of $|n| = 1 \times 10^{16}$ m$^{-2}$ for either electrons or holes. Data were taken at 1.4 K. Both on the electron and the hole side, the WAL feature can be influenced by a gate voltage, but, just as for samples A and B, the dependence does not follow the intuitive expectation.



\end{document}